\documentstyle[prb,aps,twocolumn,floats]{revtex}

\begin{document}

\title {\bf Interaction effects and the
metallic phase in p-SiGe}

\author{P.T. Coleridge, A.S. Sachrajda and P. Zawadzki \\ }

\address
{Institute for Microstructural Sciences, National Research
Council, Ottawa,
Ontario, K1A OR6, Canada\\ }

\date{3 November 2000}

\maketitle

\begin{abstract}

Magnetoresistance results are presented for p-SiGe samples on
the metallic side of the B=0 metal-insulator transition. The
results cannot be understood within the framework of standard
theories for quantum corrections of a weakly interacting 2-
dimensional system. In particular no logarithmic dependence on
temperature is observed, at low fields, in either the
longitudinal or Hall resistivities despite evidence in the
magnetoresistance of weak localisation effects. Further, the
Hall coefficient shows a strong logarithmic dependence on
field. The results are better explain by renormalisation group
theories and by an anomalous Hall effect associated with strong
spin-orbit coupling in the presence of a background spin
texture.
\vspace*{2.0cm}


\end{abstract}

%\newpage

\subsection{Introduction}

     Recent experiments indicating the existence of a metallic
state and a metal-insulator transition (MIT) in two-dimensional
(2-D) semiconductor systems \cite{} continue to attract
attention \cite{abrahams00}. There is, as yet, no general
consensus about the origin of the metallic behaviour and it
remains a controversial topic.  The metallic behaviour appears
in strongly interacting systems where r$_s$ (the ratio of the
interaction energy to the kinetic energy) is large, typically
5 to 20 and the Coulomb interaction energy is by far the
largest energy in the problem. While the existence of a MIT in
2-dimensional systems contradicts the well established one
parameter scaling theory for non- interacting systems
\cite{abrahams79}, it is not {\it a priori} forbidden for
strong interactions \cite{dobr97} and the observation of good
scaling behaviour \cite{simonian97}, with symmetry about the
critical density, supports the view that this is a genuine
transition, driven by the interactions. This is also consistent
with predictions of renormalisation group (RG) theories
\cite{abrahams99,finkelstein84,castellani84,castellani98} that
a low temperature metallic phase can exist when interactions
and disorder are both present. 

     An alternative view \cite{altshuler99} is that there is no
transition, just a cross-over from weakly localised to strongly
localised behaviour, and that the strong interactions do not
significantly modify the basic Fermi liquid character. Large
values of r$_s$ are almost inevitably associated with low
densities and Fermi energies ($k_B T_F$) that are not much
larger than the measuring temperatures so effects associated
with the small Fermi degeneracy are likely to be significant.
It is argued that these can fully account for the metallic-like
increase in resistivity with temperature and that weak
localisation and interaction effect corrections are present but
concealed by the larger semi-classical effects.

     Experimental support for this second viewpoint has
recently been presented based on results obtained in p-GaAs
\cite{simmons00} and p-SiGe \cite{senz00b}. In both cases the
B=0 resistivity shows no direct evidence of a lnT
dependence, the standard signature of both weak
localisation and interaction effects. The magnetoresistance,
however, has the characteristic negative peak associated with
the destruction, by dephasing, of a weak localisation
correction. Also the Hall coefficient shows a lnT dependence,
interpreted as evidence for interaction corrections. The
magnitude of both these terms is consistent with the standard
predictions and it is argued the behaviour is that of an
entirely conventional Fermi liquid, despite the large values of
r$_s$.

     Similar measurements, in p-SiGe samples, are presented
here \cite{ptc99}. The experimental data is consistent with
these results but more detailed and is interpreted in a rather
different fashion.  The Hall data, in particular, shows a rich
variety of field and temperature dependences that cannot be
explained by the standard theories. It is suggested these new
phenomena are associated, at least in part, with the strong
interactions and, while not explained completely,  are more
consistent with predictions of renormalisation group theories.
This supports the view that the low temperature metallic
behaviour is not that of a conventional Fermi liquid but should
be directly associated with the strong interactions.

\subsection{Theory: weak interactions}

     Because the data presented below is discussed, initially,
in terms of the quantum corrections for weakly interacting
systems it is convenient to summarise the standard theoretical
treatments of this \cite{altshuler85,fukuyama85,lee85}. The
theories start with a semi-classical description of the
transport, where the
diffusion of electrons at the Fermi level is considered to obey
classical dynamics and introduce quantum effects as small
corrections. Drude-Boltzmann theory gives the zero field
conductivity as 

\begin{equation}
     \sigma_0  =  n_s e^2 \tau/m^{\ast}
               = \frac{g_s} {2} k_f l \frac{e^2} {h}
\label{eq1}
\end{equation}
where n$_s$ is the carrier density, $\tau$ the transport
lifetime, m$^{\ast}$ the effective mass, g$_s$ the spin
degeneracy, k$_f$ the Fermi wavevector and {\it l} the mean
free path. In a magnetic field the conductivity components are
given by

\begin{equation}
\sigma_{xx}(B) = \frac{\sigma_0} {(1  +  \mu^2 B^2 )}; \;\;\;\; 
   \sigma_{xy}(B) = \mu B \sigma_{xx}(B)
\label{eq2}
\end{equation}
where the mobility $\mu$ = $e\tau / m^{\ast}$. Quantum
interference introduces a weak localisation correction to this
conductivity with a logarithmic temperature dependence 

\begin{equation}
\Delta \sigma_{xx}^{wl}(T) =  
     \alpha p  (e^2/\pi h) \ln (k_B T \tau/ \hbar)
\label{eq3x}
\end{equation}
where it is assumed the phase breaking time $\tau_{\phi}$
varies as $T^{-p}$. The amplitude $\alpha$ is expected to be 1
for normal scattering (-0.5 for pure spin-orbit scattering and
0 for spin scattering) but may also be reduced by other factors
such as anisotropic scattering.  

A similar lnT dependence results from the Coulomb
interaction effect  

\begin{equation}
\Delta \sigma_{xx}^{ee}(T) =  (1- \frac{ \textstyle 3 F^{\ast}} 
{ \textstyle 4}) (e^2/ \pi h) \ln (k_B T \tau/ \hbar) .
\label{eq4x}
\end{equation}
Two processes, with opposite signs, contribute here. An
exchange term (singlet channel), involving only small momentum
transfers, and a Hartree term (triplet channel) involving
$F^{\ast}$, the Fermi surface average of the screened Coulomb
interaction. Increasing $F^{\ast}$ implies a smaller total
correction and an increased tendency towards delocalising
behaviour. For Thomas-Fermi screening $F^{\ast} < 1$ but it
may, in principle, be larger. In this case a weakly interacting
theory is inappropriate and should be replaced by more
sophisticated approaches but with trends that are probably
still given correctly by eqn.4. Large values of $F^{\ast}$ lead
to negative coefficients for the interaction term which may
even overcome the weak localisation term and result in a total
negative, or delocalising, coefficient.  Values of $F^{\ast}$
larger than one are frequently obtained from fits to
experimental data \cite{bishop82,burdis88}.

     Application of a magnetic field allows the weak
localisation and interaction terms to be separated.  At low
fields dephasing of the weak localisation term gives a
characteristic , negative magnetoresistance,     

\begin{equation}
  \Delta \sigma_1 (B) =  \alpha \frac {e^2} {\pi h}
     [ \Psi ( \frac{1}{2}+\frac{\tau_B}{2 \tau_{\phi}})
  - \Psi ( \frac{1}{2}+\frac{\tau_B}{2 \tau})  + 
     \ln ( \frac{\tau_{\phi}}{\tau})]
\label{eq5x}
\end{equation}
where $\Psi$ is the digamma function and $\tau_B = \hbar/2eDB$
with D (the diffusion constant) = $v_F^2 /2\tau$. In terms of
the variable $h=2\tau_{\phi}/\tau_B$ (proportional to B and not
to be confused with the Planck constant) this function
varies as $h^2/24$ for small h and as ln(h) for $1 \ll h  \ll
\tau_{\phi}/\tau$.  Fitting to the characteristic shape,
particularly at small B, allows an experimental determination
of $\alpha$ and  $\tau_{\phi}$ .

     The interaction term is unaffected by small magnetic
fields but at higher fields the singlet part is modified by
Zeeman spin-splitting. This leads to a positive
magnetoresistance given by

\begin{equation}
  \Delta \sigma_2 (B)  =  - ( e^2 / \pi h) ( F^{\ast}/ 2) G(b)
\label{eq6x}
\end{equation}
where b = $g\mu_B B/ k_B T$ with g the g-factor of the spins.
The function G(b) is known: for small b  $G(b) =  0.084 b^2$
and for large b it varies as ln(b/1.3) and can be calculated in
the intervening region \cite{burdis88}.

In addition to these terms there is also the classical
magnetoconductance, eqn.2, which gives a correction with a
quadratic dependence

\begin{equation}
     \Delta \sigma_{xx}(B) \approx   - \sigma_0  \mu^2 B^2.  
\label{eq7x}
\end{equation}

     Interactions do not affect the Hall conductivity,
$\sigma_{xy}$, but the weak localisation corrections appear a
factor of two larger. When the Hall coefficient ($R_H = 
\rho_{xy} / B$ ) is obtained by inverting the conductivity
tensor the weak localisation terms therefore cancel and 

\begin{equation}
     \Delta R_H / R_H    
     =  -2  \Delta \sigma_{xx}^{ee} /\sigma_{xx}.
\label{eq8x}
\end{equation}
A logarithmic temperature dependence in the Hall coefficient is
therefore expected with an amplitude proportional to the
strength of the electron-electron interactions but independent
of weak localisation.

     At low fields, when $\rho_{xx}$ is obtained by inverting
the conductivities, the ln(T) weak localisation and interaction
effect correction terms appear additively in the same way as in
$\sigma_{xx}$ but the quadratic classical term is automatically
cancelled. At larger fields, when $\mu B$ becomes significant,
the admixture of $\sigma_{xx}$ and $\sigma_{xy}$ contributions
introduces an additional ln(T) term \cite{houghton82},
proportional to $(\mu B)^{2}$. For B sufficiently large for
this to be important the weak localisation is usually
suppressed and this quadratic term is positive with an
amplitude proportional to $(1-\frac {3}{4} F^{\ast})$.

\subsection{Samples}

     The samples used in this investigation are from a set of
p-type modulation doped strained Si-Ge quantum wells which
exhibit MIT behaviour \cite{ptc97}. They were grown using a low
temperature UHV-CVD process and have a Si buffer layer, a 40nm
Si$_{.88}$Ge$_{.12}$ quantum well, a spacer layer (of variable
thickness), a boron doped layer and a thin Si cap. The holes
reside in an approximately triangular SiGe quantum well,
produced by the asymmetric doping, which is strained because it
is lattice matched to the Si substrate. It has been established
that they are in almost pure  $|M_J|$ = 3/2   states, well
decoupled from other hole states by strain and confinement
\cite{people85}. The two samples used were Sample A  with a
density of 5.7$\times$10$^{11}$cm$^{-2}$ which is deep in the
metallic phase and sample B, with a density of
1.2$\times$10$^{11}$cm$^{-2}$, close to the critical density
for the MIT, which is at approximately
1.0$\times$10$^{11}$cm$^{-2}$. Values of r$_s$ are
approximately 4 and 6, respectively, for the two samples.  

     Effective mass values are known only approximately in this
system. Measurements from the temperature dependence of the
Shubnikov-de Haas oscillations gave values of 0.305m$_0$ for
sample A \cite{ptc96} and approximately 0.23m$_0$ for sample B.
These are significantly larger than the band mass of 0.20m$_0$
\cite{hasegawa63} and may reflect a breakdown of the standard
Lifshitz-Kosevitch expression in a system where strong quantum
corrections are expected. It should be noted that high field
cyclotron resonance measurements \cite{song95,wong95} have
given values as low as 0.18m$_0$.

     The g-factor in p-SiGe is not well known but is of order
4 in perpendicular fields \cite{glaser90}. In sample A the low
field Shubnikov-de Haas (SdH) oscillations appear initially at
odd filling factors \cite{ptc96}, corresponding to a Zeeman
splitting of the Landau levels sufficiently large that
$gm^{\ast}/m_0 > 1$. With  $m^{\ast}/m_0 \approx$ 0.25 this
means g is probably slightly larger than 4. In sample B the SdH
oscillations appear initially at even filling factors so
$gm^{\ast}/m_0 < 1$ . Using m$^{\ast}$ = 0.23m$_0$  and taking
$gm^{\ast}/m_0$ to be 0.8$\pm$ 0.2 gives g = 3.6 $\pm$ 1. In
tilted fields, the almost pure  $|M_J|$ = 3/2 character of the
heavy hole states means the g-factor varies accurately as
Bcos$\theta$ \cite{ptc96}, where $\theta$ is the tilt angle
away from the perpendicular. Therefore, unlike the situation in
Si-MOSFETs and n-GaAs heterojunctions, tilting the magnetic
field leaves the relative contributions from orbital and spin
splitting terms unchanged but a function of Bcos$\theta$.

\subsection{Experimental procedures}

     Measurements were made using a DC bridge \cite{avs}, with
current reversal at 15Hz. Two cryostats were used, a dilution
refrigerator giving temperatures below 100mK and a sorption
pumped He3 system that could achieve sample temperatures down
to 270mK. In both cases thermometry was in a field free region,
with the sample cooled essentially by the copper leads.
Measurements using a calibrated, field insensitive, thermometer
in place of the sample indicated the temperature gradients were
less than 5mK.

     Resistance measurements, especially in sample B, showed an
unusual sensitivity to measuring current at the lowest
temperatures. Currents as low as 1nA, corresponding to voltages
of only a few microvolts, produced detectable metallic-like 
non-linearities in the I-V characteristics and made accurate
low temperature measurements of the zero field resistivity
difficult. This is attributed, at least to some extent, to the
proximity of the MIT. Measurements in magnetic fields,
especially near B=0, were complicated by an extreme sensitivity
to sweep rate and long thermal time constants. This is not
completely understood although it is accounted for, in part, by
heating associated with flux jumps as field in the
superconducting magnet was reversed. Care was taken to average
results from up and down sweeps and to check frequently (making
corrections if necessary) that the measured resistances were
the same in static and swept fields. 

\subsection{Experimental results} 

     Magnetoresistance data from sample A is shown in figure 1.
At zero field the monotonic metallic-like increase of
resistance with temperature, characteristic of samples showing
a MIT, can be clearly seen. Over the temperature range shown
(0.15 - 2.2K) a logarithmic weak localisation term (eqn.3)
would give a {\it decrease} of resistance with increasing
temperature of order 50-100 ohms/square, roughly a factor of
two larger and in the opposite sense to the observed variation.
The well defined low field negative magnetoresistance, with a
width that increases with temperature, is attributed to weak
localisation and the positive magnetoresistance at higher
fields to a Zeeman interaction term. 

     Attempts to fit the field dependence to a combination of
eqns. 5 and 6 failed, even when the full field dependence of
G(b) was used,  with the g-factor as a separate fitting
parameter for a total of five.  The alternative approach taken
was therefore to fit the data to eqn. 5 at the lowest fields,
using just three parameters: $\tau$ (determined essentially by
the B=0 value of the resistivity), $\tau_{\phi}$ and an
amplitude $\alpha$. The results of these fits are shown in
figure 1. Care was taken to establish that the parameters were
insensitive to the precise range of B and also that they were
not significantly altered when this range was extended and an
extra quadratic term added. The values of $\tau_{\phi}$,  shown
in figure 2, are similar to those obtained from other
measurements in p-SiGe \cite{emeleus93,cheung94,senz00a}.
Values of $\alpha$ were 0.7 at low temperatures decreasing to
0.6 at higher T.

      The residues from these fits are shown in fig.3a. Plotted
against B/T (fig. 3b) they collapse onto a single curve which
gives some confidence that the fitting procedure has
successfully separated the weak localisation term from a Zeeman
interaction term which is a function of B/T. This curve is not,
however, given by eqn.6. The solid line in fig. 3b is G(b),
fitted to the curve for larger values of B/T, with $F^{\ast}$
= 2.45 and a g-factor of 6.4. It deviates significantly from
the data at low values of B/T and the value for g is somewhat
larger than expected. 

     The most important conclusion from this fitting procedure,
despite only qualitative agreement between theory and
experiment, is the large value of $F^{\ast}$. This implies that
the interactions are strong and that the standard theory, where
it is implicitly assumed that  $F^{\ast}\leq 1$, is inadequate.
Expressions for the Zeeman interaction term using
renormalisation group theory have been derived, for b $\gg$ 1,
by Finkel'stein \cite{finkelstein84}

\begin{equation}
  \Delta \sigma_2 (B)  =  - ( e^2 / \pi h) \, 2 \, 
  [\frac{1 + \gamma_2}{\gamma_2} \ln(1 + \gamma_2) - 1 ]      
 \ln ( b )
\label{eq9x}
\end{equation}
and for b $\ll$ 1 by Castellani {\it et al } 
\cite{castellani84,castellani98} 

\begin{equation}
  \Delta \sigma_2 (B)  =  - 0.084 ( e^2 / \pi h) \gamma_2  (1
+ \gamma_2) b^2 .
\label{eq10x}
\end{equation}
Here the coupling constant $\gamma_2$ is a measure of the
renormalised interaction strength for the triplet scattering
channel. It reduces to F$^{\ast}/2$ in the limit of weak
interactions.

     The dashed lines in fig. 3b show fits using these two
equations in the high field and low field limits respectively,
using g = 3.6 and $\gamma_2$ = 2.6 in both cases. Making a
reasonable interpolation between the high field and low field
expressions this gives a much better description of the
experimental data. The value of g is perhaps a little small but
is predominantly determined by the low field quadratic regime
where experimental errors are relatively large.

     For sample B the magnetoresistance data (figure 4) is
shown as the conductivity obtained by inverting measured values
of $\rho_{xx}$ and $\rho_{xy}$. Again the increase in
conductivity with decreasing temperature is in the opposite
sense to that expected for a weak localisation lnT term which,
over the temperature range shown, should be of order 0.5
e$^2$/h. Because the ratio  $\tau_{\phi}/\tau$ is significantly
smaller than in sample A, the weak localisation peak is only
poorly defined and the behaviour is dominated by a quadratic
dependence on field which is a combination of the classical
term (eqn. 7) and the low field Zeeman interaction term (eqns.6
or 10). Because the g-factor varies quite precisely as
cos$\theta$ these results, when plotted against Bcos$\theta$,
are essentially independent of tilt. Spurious sweep rate
effects associated with heating (noted above) are dependent on
the total field and are less apparent in tilted fields so it
was convenient to use tilted field data (at an angle of 69$^o$)
for a detailed analysis of the field and temperature
dependences. 

     The structure at low fields is sufficiently weak that the
fitting procedure developed for sample A could not be used: the
equations became ill-conditioned and gave imprecise and
interdependent values of $\alpha$ and $\tau_{\phi}$. The fit is
relatively insensitive to the exact value of $\alpha$ (see also
Senz {\it et al} \cite{senz00a}) so this was fixed (at 0.65)
and just three adjustable parameters used: $\tau$,
$\tau_{\phi}$ and a quadratic coefficient. As can be seen in
figure 4 these gave very good descriptions of the data provided
the fits were restricted to the field region where the Zeeman
interaction term is expected to be quadratic in B. Values of
$\tau_{\phi}$ are shown in figure 2.

     When the weak localisation and classical mobility terms
are subtracted off, the residues should be the Zeeman
interaction term. This is plotted against B/T in figure 5. The
collapse onto a single curve is not as good as in sample A,
mainly because the classical mobility correction is relatively
large (especially at higher temperatures) and there is some
uncertainty about how  $\sigma_{0}$ should be defined
\cite{footnote1}. Using a g-factor of 3.6 the quadratic fit
shown gives  $\gamma_2$ = 0.7 $\pm$ 0.2 if eqn. 10 is used or 
$F^{\ast}$ = 2.5 using eqn.6.  In this case the deviation away
from quadratic behaviour ar higher fields is better described
by eqn. 6 with  $F^{\ast}$ = 2.5 than by the RG eqn.9 with
$\gamma_2$ = 0.7.

     In both samples the amplitude of the Zeeman interaction
term gives large values of $F^{\ast}$ (or of $\gamma_2$)
consistent with strong Coulomb interactions. Another measure of 
$F^{\ast}$, at least according to the standard theory, is
obtained by looking at the magnitude of the lnT term in the
Hall coefficient (eqns.4 and 8). In practice, as shown in
figure 6 where $R_H$ is plotted against field, the situation is
more complicated. In both samples $R_H$ is
strongly field dependent and at low fields, where the lnT term
is expected, there is no detectable temperature
dependence. The absence of a low field lnT term (see also the
inset to figure 6b \cite{footnote2}) in $R_H$ parallels the
absence of a lnT dependence in the B=0 conductivity data. For
the conductivity alone this might perhaps reflect a fortuitous
cancellation between weak localisation and interaction terms of
the opposite sign but the absence of any T dependence in the
Hall coefficient, which is affected only by interaction
corrections, shows that the lnT weak localisation and
interaction terms must separately vanish despite the evidence
from the magnetoresistance of weak localisation. There is no
explanation for this within the framework of the ``standard''
theory for quantum corrections. Similarly there is no
explanation for the strong field dependence of $R_H$ near B=0.

     At higher fields this field dependence persists and
temperature dependence also eventually appears. A semi-log plot
of the data, figure 7, shows that once developed this is,
indeed, linear in lnT. For sample A the slope, 
 $\Delta R_H /(R_H \ln T)$,  is -0.024, independent of field.
For sample B it varies with field with a maximum value (at 0.5
tesla) of -0.085. Expressed in terms of an equivalent amplitude
for $\Delta \sigma_{xx}^{ee}$ (eqns.4 and 8) these correspond
respectively to $F^{\ast}$ = 0.6 and 0.74. The lnT term at
higher fields is therefore consistent with other measurements
in p-SiGe \cite{emeleus93,senz00a} where values of $F^{\ast}$,
determined in the same way, were found to be about 0.85. In
these measurements no special attention was paid to the very
low field region and non-linearities in $R_H$ were not
reported.  

     The lnT term developes in a different way in the two
samples. In sample A there is a fairly abrupt transition from
a high T regime, where $R_H$ is temperature independent, to a
low T limit where the slope appears to be independent of B. The
breakpoint where this change of slope occurs is found
empirically to depend linearly on B with (B/T)$_{\rm break}
\approx $ 0.9 tesla/kelvin. In sample B it develops at lower
fields and, as shown in the inset to figure 7, the slope
increases linearly with field with a low field extrapolation
corresponding to a vanishing slope at B=0. In sample B
measurements of $R_H$ at higher fields are complicated by the
onset of the quantum oscillations which, because of their long
period, cannot be separated from the background variation.

     As previously noted the B-dependence of $R_H$ is not
understood.  Semi-log plots of the data (figure 8) show that at
low fields $R_H$ increases approximately as lnB and at higher
fields, when the temperature dependence appears, the increase
remains predominantly logarithmic but with a (T dependent)
reduction in the slope. The low field slopes, given by $\Delta
R_H/(R_H \ln B )$, are 0.11 and 0.18 for samples A and B
respectively. If attributed to a coulomb interaction term they 
correspond to conductivity corrections (using eqn.8) of
$\Delta\sigma_{xx}^{ee} = -2.4 (e^2/\pi h)\ln(B)$  and $-0.8
(e^2/\pi h)\ln(B)$. These are relatively large effects,
particularly for sample A, and should be readily observed in
the direct measurements of $\sigma_{xx}$ (or $\rho_{xx}$).  It
should be noted that expressed in terms of a conductivity
correction, the amplitude of the B dependence in $R_H$ is
larger for the higher density sample, deeper into the metallic
phase. Indeed, for both the samples it is, probably
fortuitously, equal to the value of $\gamma_2$ derived by
fitting the Zeeman interaction term.

\subsection{Discussion}

     The data presented above cannot be understood within the
framework of the standard theory for weakly interacting systems
outlined above. One problem is a magnetoresistance consistent
with weak localisation effects but without any sign of the
corresponding lnT dependence in $\rho_{xx}$.  A second problem
is the large values of $F^{\ast}$ extracted by fitting the
Zeeman Interaction term without, again, any corresponding zero
field lnT dependence. Thirdly, there is the strong lnB
dependence of $R_H$, particularly near B=0. 

     The weak localisation peak in the low field
magnetoresistance appears conventional. It can be fitted by the
standard expression with values of $\tau_{\phi}$ of the
expected order of magnitude. If the dephasing occurs by
inelastic carrier-carrier scattering in the limit of small
momentum transfer values (sometimes known as Nyquist dephasing)
$\tau_{\phi}$ should be given by  
\cite{altshuler85,brunthaler00,altshuler82} 

\begin{equation}
     \tau_{\phi} = \hbar f(g)/k_B T  
\label{eq11x}
\end{equation}
where f(g), with the conductance g in units of $e^2/h$, is $g/
\ln(g/2)$ for $g \gg 1$ but of order g as g approaches 1.
Figure 2 shows these values plotted with f(g) = 7.4 for sample
A (where g=15) and 2.6 for sample B (where g=2.6). The
experimental values of $\tau_{\phi}$ are approximately 5 times
smaller than this; similar discrepancies are also seen in Si-
MOSFETs \cite{brunthaler00} and in p-GaAs \cite{simmons00}.  It
is not clear whether this discrepancy is significant.

     Interpreted according to the standard theory the Zeeman
interaction term is very large, with values of F$^{\ast}$
significantly larger than one. Similar discrepancies exist in
Si-MOSFETs \cite{bishop82,burdis88}. Also the functional
dependence on B/T appears to be incorrect. A plausible
extension of the weakly interacting theory, with large values
of F$^{\ast}$, might be able to explain the absence of a lnT
term in $\sigma_{xx}$ as a cancellation between a positive
coefficient for the weak localisation correction and negative
values of the interaction coefficient so  $(\alpha p + 1-
\frac{3}{4} F^{\ast})$ vanishes. But $R_H$ should then also
have a lnT term, dependent only on the interactions, with a
coefficient $(1- \frac{3}{4} F^{\ast})$.  With  $\alpha p$
(obtained from the magnetoresistance) of order 0.6 these
coefficients cannot vanish simultaneously. Therefore, even if
empirically adjusted to allow for large values of $F^{\ast}$,
the standard theory cannot account for the absence of a ln(T)
term in both $\sigma_{xx}$ and $R_H$.

     At low fields, according to the standard theories,
interactions do not give a lnB dependence in the Hall
coefficient and while the weak localisation function (eqn. 5)
has a lnB dependence it is of the opposite sign and appears
only for $1 \ll 2\tau_{\phi}/\tau_B \ll \tau_{\phi}/\tau$. In
sample A, for example, this corresponds to fields less than .05
tesla and is a factor 3 or 4 times smaller than the term
corresponding to the observed $R_H$ dependence. The absence of
a ln(B) term in $\sigma_{xx}$ matching that in $R_H$ cannot
therefore be attributed to a fortuitous cancellation with a
weak localisation term. 

     At high fields the Zeeman Interaction term has a ln(B)
dependence. For, sample A, this corresponds to that measured in 
$R_H$ to accuracy of about 20\%. For sample B, where the ln(B)
dependence in $\sigma_{xx}$ could not be explicitly determined,
the $R_H$ value is also qualitatively consistent with eqn.9 and
a fitted value of $\gamma_2$ = 0.7. In both cases, at higher
fields, the {\it field} dependence of $R_H$ can therefore
probably be accounted for by the Zeeman interaction term but
not the {\it temperature} dependence. The B/T dependence seen
in the $\sigma_{xx}$ data is not reproduced in the $R_H$ data. 

     Some of these results can be understood within the
framework of renormalisation group (RG) theories
\cite{finkelstein84,castellani84,castellani98}. These involve
a coupling constant $\gamma_2$, related to the strength of spin
density fluctuations. For weak interactions $\gamma_2$ reduces
to $F^{\ast}/2$ but for strong interactions it may be
dramatically enhanced by correlations. Disorder leads to a
renormalisation of the interaction amplitudes and for weak
disorder $\gamma_2$ is expected to increase with decreasing
temperature. The localising lnT dependence is then replaced by
scaling towards metallic behaviour with, in the T=0 limit, an
eventual saturation at a finite resistivity.  For strong
disorder scaling is towards an insulating state so there should
be a cross over from metallic to insulating behaviour as the
disorder increases. In terms of the dimensionless
resistivity r (in units of $\pi h/e^2$) the equations diverge
when $r\gamma_2 \sim 1$ and the theory is unable to answer the
question of whether a quantum critical point is expected at the
cross-over.

     RG theories account for the saturation of the low
temperature resistivity at a finite value and the absence of a
lnT term at B=0 (although they do not predict the exponential
dependence on T that is frequently observed). The large values
of $F^{\ast}$ (or equivalently $\gamma_2$) needed to describe
the magnitude of the Zeeman Interaction term are expected, also
the observation that $\gamma_2$ seems to be larger in the
higher density sample, further away from the critical density.
The experimentally observed functional dependence of the Zeeman
term also seems to be in better agreement with RG expressions
than with G(b).  One problem is that renormalisation implies
$\gamma_2$ should increase with decreasing temperature in which
case the functional dependence on B/T seen in figure 3b should
be destroyed. This can probably be explained by the fact that
most of the data is restricted to the low temperature region
when the resistivity is close to saturation and $\gamma_2$ is
no longer temperature dependent.

     A further prediction of RG theory \cite{castellani84} is
that in the high field limit $\Delta \sigma_{xx}(T) = \beta
(e^2/\pi h) \ln(T) $ where $\beta$ has the universal value,
2-2ln2 (= 0.61). High field in this case corresponds to a
Zeeman splitting g$\mu_B$B $\gg$ k$_B$T, ie $b \gg 1$. This
universal dependence is not seen directly in $\sigma_{xx}$ (it
would correspond for example to a value of $\gamma_2 \approx$
0.7 in eqn. 9 for the Zeeman Interaction term) but a universal
value of $\beta$ = 0.6 does seem describe the temperature
dependence of the high field, low temperature data for $R_H$ in
sample A. The cross-over to this behaviour, at B/T $\approx$
0.9 tesla/kelvin, corresponds to b $\approx $ 2. That sample B
does not behave in the same way may be associated with the fact
that in this sample, close to the MIT,  r$\gamma_2$ is of order
0.8 and in the regime when the RG equations are starting to
fail.

     The behaviour that cannot be understood, according to
either the weakly interacting or RG theories, is the low field
lnB dependence of $R_H$ and the absence of a corresponding term
in  $\sigma_{xx}$. It would therefore appear that this is not
a conventional interaction term but involves some other
mechanism. Against this the correlation between the magnitude
of the lnB dependence and density and the smooth transition to
a high field behaviour which can be at least partly understood
in terms of RG theory, suggests it is nevertheless related, in
some way, to the strong interactions.

     One clue is the absence of such behaviour in Si-MOSFETs
where it might also be expected. Recent experiments in tilted
fields \cite{vitkalov00a}, in good agreement with results in
the literature for perpendicular fields, find $R_H$ increases
by less than 6\% for in-plane fields up to 12T (perpendicular
fields up to 0.5T). An essential difference between the two
systems is the strong spin-orbit coupling in p-SiGe. This
points to the field dependence of $R_H$ as being a
manifestation of the Anomalous Hall effect \cite{hurd72}. 

     The Anomalous (or Extraordinary) Hall effect is best known
in metallic ferromagnets where the Hall coefficient has a
strong field dependence given by

\begin{equation}
  \rho_{xy}    =    R_H B  + R_s \mu_0 M
\label{eq12x}
\end{equation}
with M the magnetisation and where R$_s$ is often significantly
larger than R$_H$.  This is not well understood theoretically.
According to conventional understanding, it can be explained in
ferromagnets in terms of interference between spin-orbit
coupling and spin-flip scattering \cite{karplus54,maranzana67}.
An important feature of the theories is the requirement that
the spin-up and spin-down populations are different and
proportional to the magnetisation. The effect is also observed
in magnetic semiconductors \cite{ohno94} and in materials
having a ``colossal magnetoresistance''\cite{chun00}. A recent
theory \cite{ye99} explains the results in this last example in
terms of the effect of spin-orbit coupling on the ``Berry
phase'' associated with a spin texture. It seems unlikely that
a single theory can explain all occurences of the anomalous
Hall effect but a feature of all theoretical descriptions is
the importance of the spin-orbit interaction and a
magnetisation associated with some form of spin polarisation.

     In Si-MOSFETs experiments in tilted fields \cite{CUNY}
show clearly that the spins are intimately connected with the
MIT.  For example, polarising the spins drives the system from
a metallic state towards the insulating phase and there is
evidence for a quantum phase transition into a
ferromagnetically polarised ground state at the critical
density \cite{vitkalov00b}. In p-SiGe similar experiments
\cite{unpub,senz98} show effects in the same sense but
requiring fields at least an order of magnitude larger. This is
to be expected, because the spins cannot be decoupled from the
orbital motion and fields in the quantum limit, ie with a
Landau level filling factors of order one, are required to
produce a spin polarisation \cite{footnote3}. There seems to be
no fundamental reason to suppose that the spins do not play the
same kind of role in p-SiGe as in MOSFETS, just that the strong
spin-orbit coupling makes it difficult to probe this state by
using tilted fields. The spin polarisation is, however,
revealed indirectly through the anomalous Hall effect. It would
be interesting to see if p-GaAs, where the spin-orbit coupling
is also strong, shows similar effects.

    Senz {\it et al} \cite{senz00b} have suggested that a
conventional lnT weak localisation term is in fact present in
p-SiGe and that the measured resistivity is the sum of this and
a linear term associated with the temperature dependence of the
dielectric screening function. These combine to give an  {\it
apparent} low temperature saturation of resistivity. In a
similar analysis for the data reported here figure 8 shows both
the measured conductivity and the conductivity with a putative
weak-localisation correction, $\alpha p (e^2/\pi h) \ln(k_B T
\tau/\hbar)$ subtracted off. The expected variation associated
with temperature dependent screening, is given by the
calculation of Gold and Dolgopolov \cite{gold85} as

\begin{equation}
     \sigma(T)   
          = \sigma(0) [1 - C_p  \frac{T}{T_F} + 
O ( \frac{T}{T_F})^{3/2}]. 
\label{eq13x}
\end{equation}
Similar results have also been obtained numerically by Das
Sankar \cite{dassarma86}. 

While the corrected data appears to have a predominantly linear
dependence there is, in both samples, indications of an upturn, 
larger than the experimental errors, at low T. This upturn,
which reflects directly the lnT dependence of the correction
term added to either the saturation or approximately linear
variation in the measured data, tends therefore to contradict
the conclusions of Senz {\it et al}. Lower temperature data is,
however, needed to decide this.   The coefficients C$_p$ are
larger than expected. With the impurity scattering and
interface roughness scattering dominating C$_p$ for sample A is
expected \cite{plews97} to be 1.2 - 1.5,  compared with an
experimental value of 2.5 and in sample B the measured value of
5.9 is three times the expected value of about 2.  

     Hamilton {\it al} \cite{hamilton00} draw attention to
similarities in the T dependence of p-GaAs and Si-MOSFETs. For
large values of T/T$_F$, up to about 0.4, they find an
approximately universal behaviour of $\Delta \sigma / \sigma$,
consistent for example with temperature dependent screening.
There is, however, little data for the range T $<$ .02T$_F$
where any ln(T) dependence of interaction and weak localisation
effects will be readily separable from semi-classical
behaviour. Therefore, although it seems clear that temperature
dependent screening (or similar semi-classical effects) are
important at higher temperatures it seems unlikely that they
can account for the absence of any observable lnT weak
localisation term at zero field.

\subsection{Conclusions}

     Transport data is presented for two p-SiGe samples on the
metallic side of the MIT. The metallic behaviour and the large
Zeeman interaction term cannot be understood within the theory
of conventional weakly interaction quantum corrections but are
more consistent with predictions of Renormalisation Group
theories. Fitting to RG expressions gives values for the
parameter $\gamma_2$ that are large and increase with
increasing
density.

      The low field magnetoresistance shows evidence of the 
destruction, by dephasing, of a weak localisation term but
there is no clear evidence of the corresponding lnT dependence,
localising or delocalising, in either the conductivity
$\sigma_{xx}$ or in the Hall coefficient. Phase breaking times
extracted from fits to the magnetoresistance are somewhat
smaller than expected but of the expected order of magnitude.
At higher fields, when the Zeeman splitting is larger than the
temperature, a lnT dependence develops with some evidence, in
the Hall coefficient,  for the ``universal'' behaviour
predicted, in this limit, by RG theory.

     A strong and unexplained lnB dependence in the Hall
coefficient is not reflected in the low field conductivity. 
This is tentatively associated with an anomalous Hall
coefficient resulting from the large spin-orbit coupling in
this system. If this interpretation is correct it demonstrates
directly the development a spin polarisation similar to that
reported for Si-MOSFETs and closely related to the strong spin
density fluctuations predicted by RG theories. 

     This accumulated  evidence supports the view that the
metallic behaviour and MIT in p-SiGe cannot be adequately
described by the standard theory for weak coulomb interactions
but rather should be directly associated with the strong
interactions.

\subsection{Acknowledgements}

     H. Lafontaine and R. Williams are thanked for growth of
the samples, Y. Feng and J. Lapointe for sample preparation and
R. Dudek for technical assistance. Hospitality at the Aspen
Center for Physics, where part of this work was written up, is
acknowledged by PTC.

\clearpage

\begin{figure} [t]  
\vspace*{7.5cm}  
\includegraphics{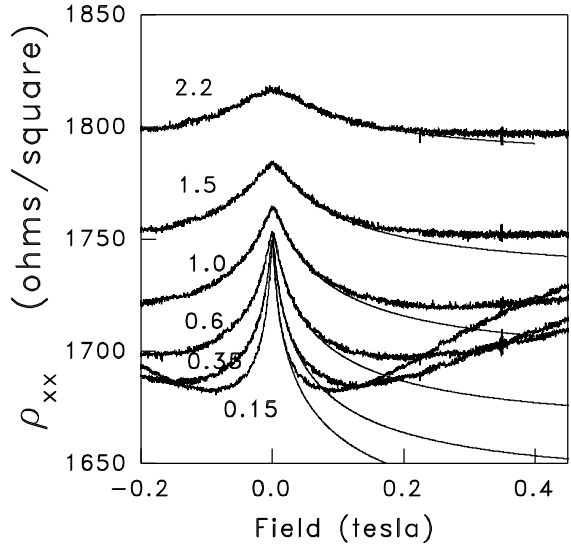}  
\caption  
{Magnetoresistance data for sample A at temperatures of 0.15, 
0.35,0.6, 1.0, 1.5 and 2.2 K (top curve).   Lines are fitted to
the weak localisation peak (eqn.5) in the low field regime as
described in the text. 
}
\label{fig1}  
\end{figure}

\begin{figure} [ht]  
\vspace*{6.5cm}  
\includegraphics{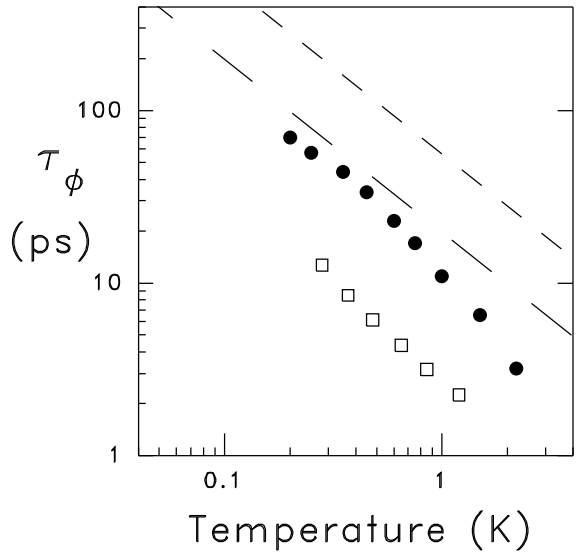}   
\caption  
{Dephasing times,  $\tau_{\phi}$, extracted from fits to the
weak localisation peak in sample A (solid points) and sample B
(open points). The lines are the values given by eqn.11 with
f(g) = 7.4 (sample A, short dashes) and 2.6 (sample B, long
dashes). For comparison $\tau$, the transport lifetime is
approximately 0.9ps in both samples.
}
\label{fig2}  
\end{figure} 

\clearpage
%\newpage

\widetext

\begin{figure} [t]
\vspace*{8.5cm}  
\includegraphics{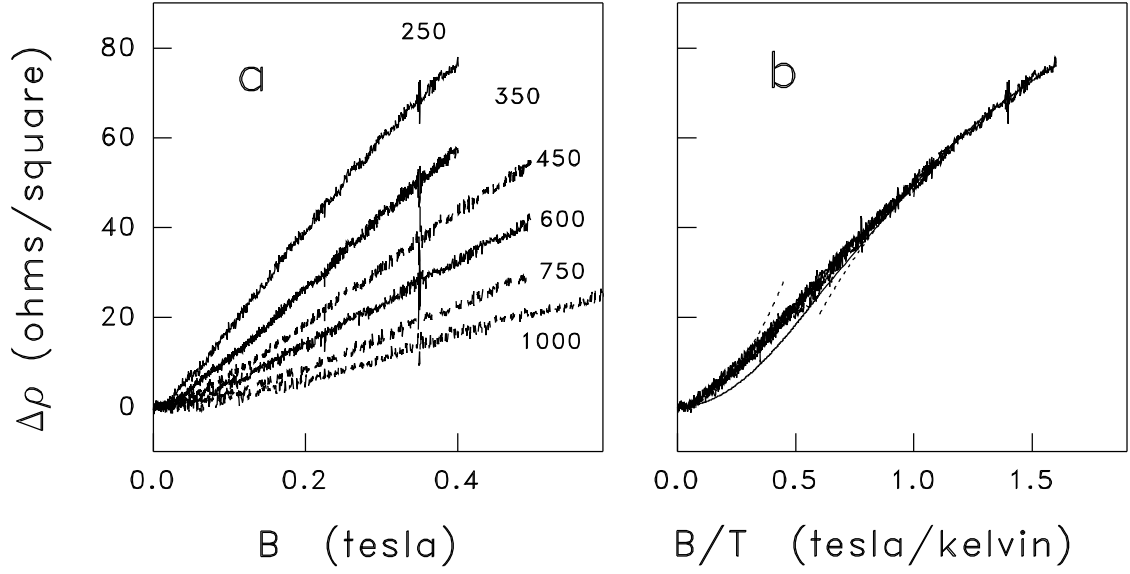}   
\caption  
{(a) Residues from the weak localisation fits to the low field
magnetoresistance for sample A. Temperatures are
respectively 250 (largest resistivity values),350, 450, 600,
750 and 1000 mK. (b) Data plotted against B/T. Solid line is a
fit to G(b) (eqn.6) and the dashed lines fits to eqns. 9 and
10. For b=1, B/T is approximately 0.4 tesla/kelvin.
}
  
\label{fig3}  
\end{figure}  
  
\narrowtext

\begin{figure} [ht]  
\vspace{8.0cm}  
\includegraphics{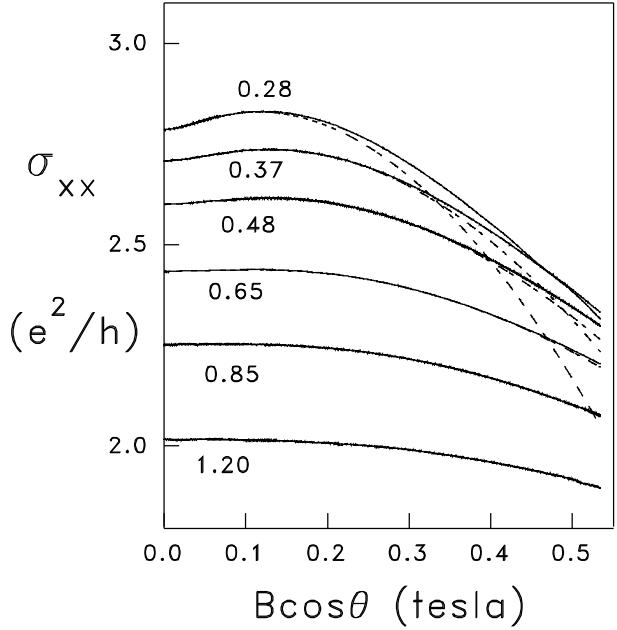}   
\caption  
{Magnetic field dependence of the conductivity in sample B at
temperatures of 0.28 (largest values), 0.37, 0.48, 0.65, 0.85
and 1.20K. As explained in the text the field is tilted at an
angle of 69$^o$ from the perpendicular. Dashed lines are fits
to eqn.5 plus a quadratic term for Bcos$\theta/T \leq$ 0.4
tesla/kelvin.  }
\label{fig4}  
\end{figure}  

\newpage

\clearpage
  
\begin{figure} [t]
\vspace{7.5cm}
\includegraphics{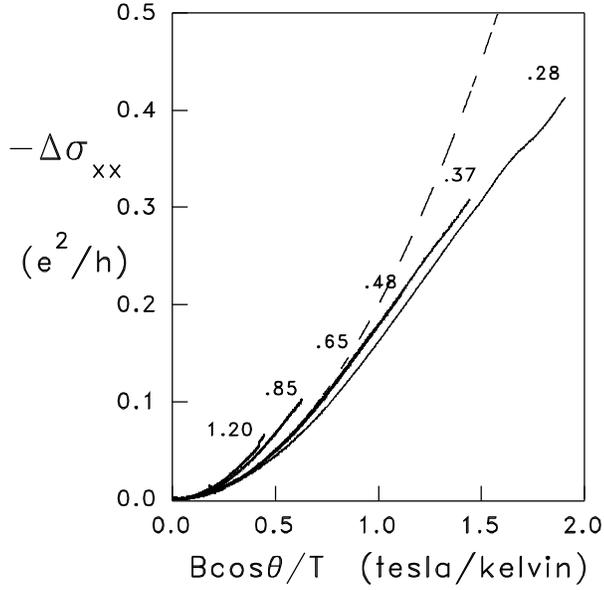}   
\caption  
{Residues from data in figure 4 with weak localisation(eqn.5)
and classical (eqn.7) terms subtracted off, plotted as a
function of Bcos$\theta$ /T. Note that data for 0.37, 0.48 and
0.65K overlap and cannot be distinguished. Dashed line is
quadratic fit used (with eqn.10) to obtain a value for
$\gamma_2$.
}  
\label{fig5}  
\end{figure}

\widetext

\begin{figure} [ht]
\vspace{8.0cm}
\includegraphics{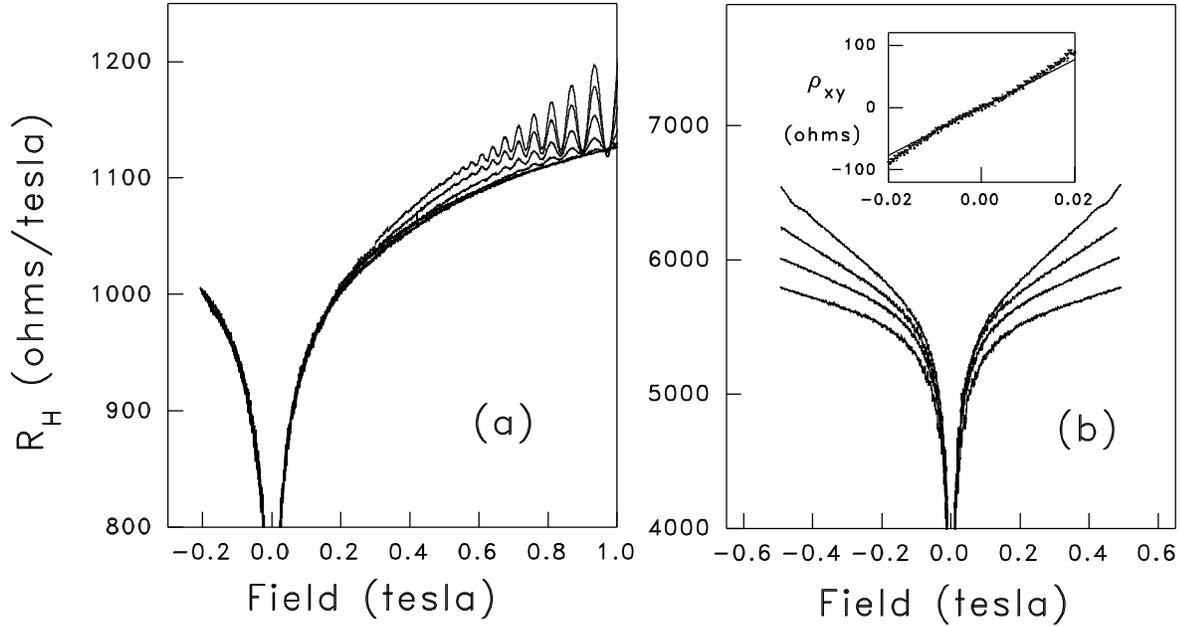}   
\caption  
{Hall coefficient as a function of field. (a) Sample A at
temperatures of 0.27, 0.39, 0.48, 0.85, 1.20 and 1.89K ($R_H$
decreases with increasing temperature). (b) Sample B at
temperatures of 0.28, 0.45, 0.75 and 1.20K. Inset shows the
very low field $\rho_{xy}$ data with a straight line
corresponding to $R_H$ = 3850 ohms/tesla. The periodicity of
the Shubnikov-de Haas oscillations corresponds to a Hall
coefficient of 1085 ohms/tesla for sample A, and 5000
ohms/tesla for sample B.
}  
\label{fig6}  
\end{figure}  
   
\narrowtext

\clearpage

\begin{figure} [t]
\vspace{12.5cm}
\includegraphics{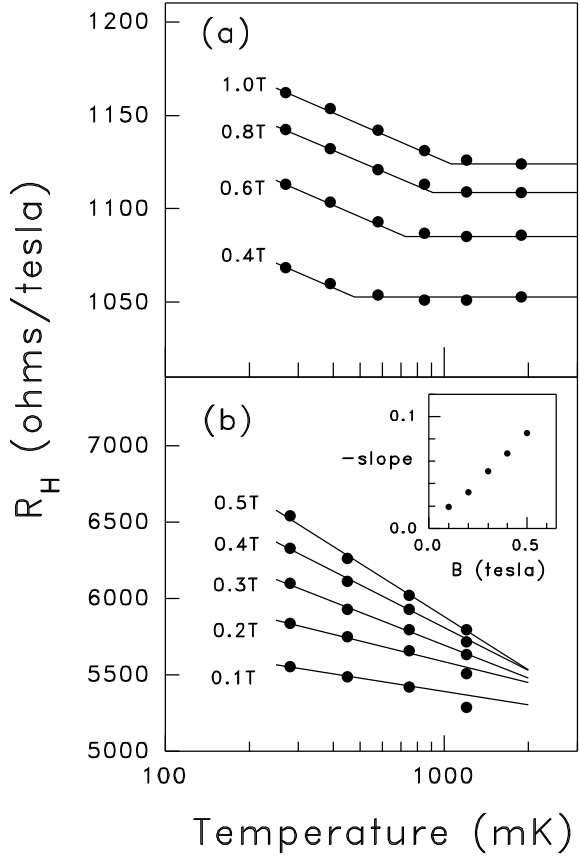}   
\caption  
{Hall coefficient as a function of temperature. (a) Sample A at
fields of 0.4, 0.6,.0.8 and 1.0 tesla. Lines are a ``guide to
the eye'' showing the transition between temperature
independent values at high T and a constant slope at lower T.
(b) Sample B at fields of 0.1 - 0.5 tesla. The slopes indicated
by the solid lines are plotted in the inset.
}  
\label{fig7}  
\end{figure}

%\newpage  
%\clearpage

\begin{figure} [ht]
\vspace{11.0cm}
\includegraphics{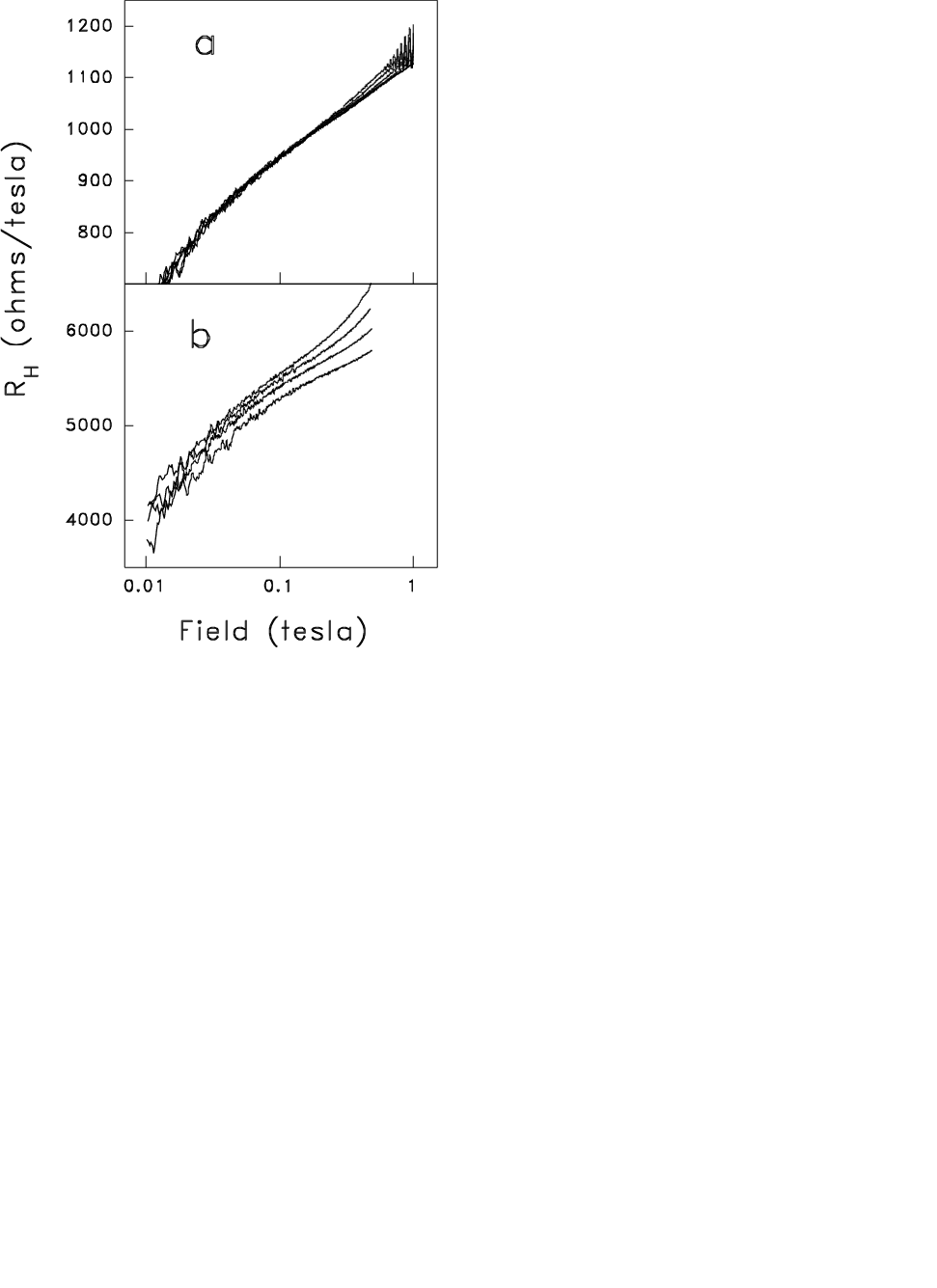}   
\caption  
{Semi-logarithmic plot of the Hall coefficient as a function of
field. Temperatures as in figure 6. (a) Sample A; (b) Sample B.
}  
\label{fig8}  
\end{figure}  

\clearpage

\begin{figure} [t]
\vspace{10.0cm}
\includegraphics{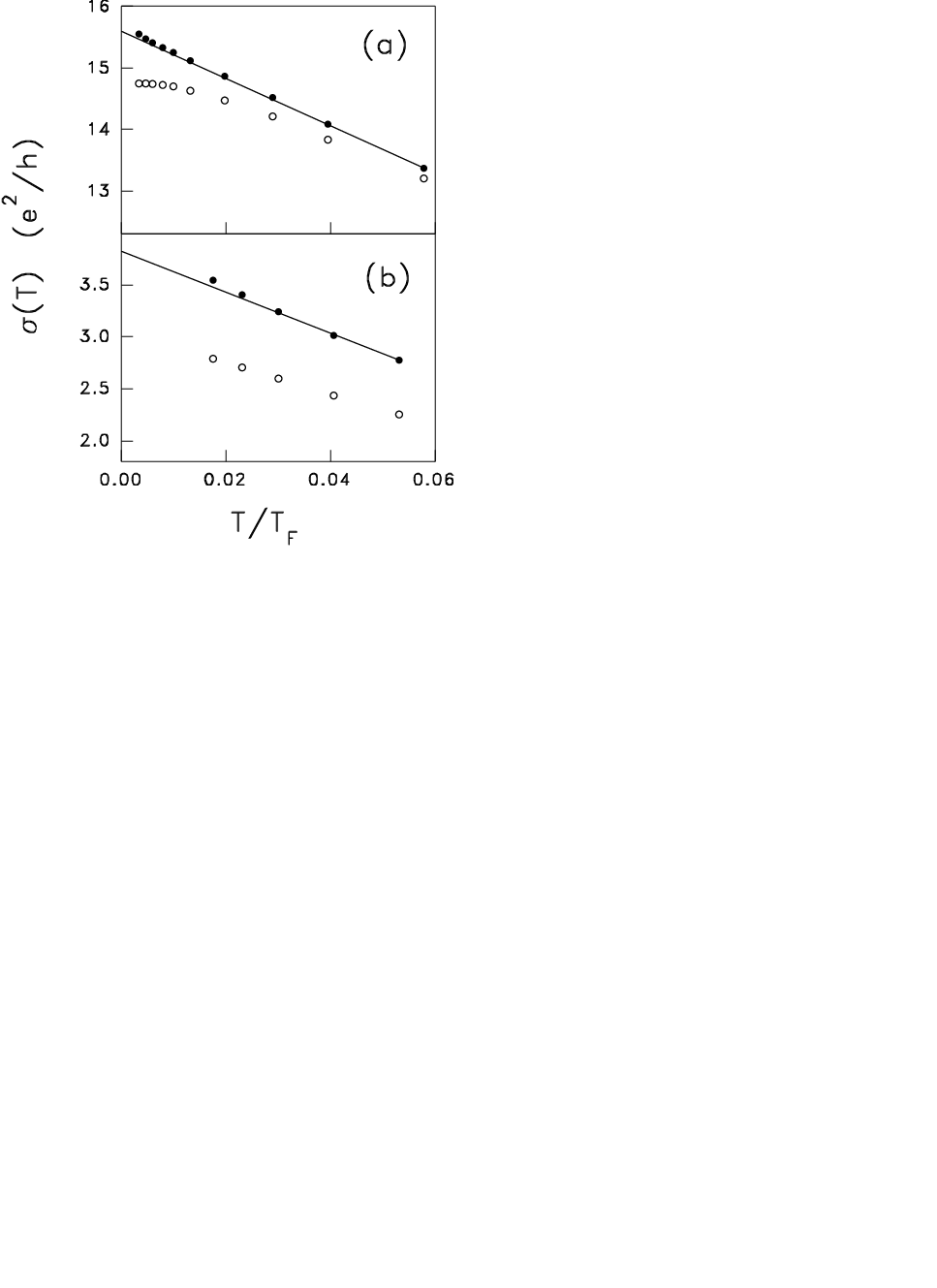}   
\caption  
{Zero field conductivity plotted against T/T$_F$. (a) Sample A
with T$_F$ = 76K. (b) Sample B with T$_F$ = 16K. In each case
the open points are the measured data and the solid points have
a ln(T) weak localisation correction subtracted. Solid lines
are linear fits to the high T values.
}
\label{fig9}  
\end{figure}  


\begin{references}
\bibitem{footnote1}    The procedure used was to take the zero
field value for $\sigma_0$ and a mobility $\mu$ determined by
a quadratic fit to $\rho_{xx}^2 + \rho_{xy}^2$.
\bibitem{footnote2}  Note that near B=0 the Hall resistivity
showed a small hysteresis and that the data presented is the
average of sweeps in both directions.
\end{references}
\end{document}